\documentclass[a4paper,12pt]{article}
\usepackage[greek,english]{babel}
\usepackage[iso-8859-7]{inputenc}
\usepackage{amsmath}
\usepackage{amsfonts}
\usepackage{amsfonts}
\begin{document}

\title{{\bf Classical and Quantum Bianchi Type III vacuum Ho\v rava - Lifshitz Cosmology}}
\vspace{1cm}
\author{\textbf{T. Christodoulakis}\thanks{tchris@phys.uoa.gr}\,,~\textbf{and} \textbf{N. Dimakis}\thanks{nsdimakis@gmail.com}\\
University of Athens, Physics Department\\
Nuclear \& Particle Physics Section\\
Panepistimioupolis, Ilisia GR 157--71, Athens, Hellas}
\date{}
\maketitle
\begin{center}
\textit{}
\end{center}
\vspace{-1cm}
\numberwithin{equation}{section}
%%%%%%%%%%%%%%%%%%%%%%%%%%%%%%%%%%%%%%%%%%%%%%%%%%%%%%%%%%%%%%%%%%%%%%%%%%%%%%%%%%%%%%%%
\begin{abstract}
A diagonal Bianchi Type III space-time is treated, both at the classical and quantum level, in the context of {\it Ho\v rava - Lifshitz} gravity. The system of the classical equations of motion is reduced to one independent Abel's equation of the first kind. Closed form solution are presented for various values of the coupling constants appearing in the action. Due to the method used, solutions of Euclidean, Lorentzian and neutral signature are attained. The solutions corresponding to $\lambda\neq1$ are seen to develop curvature singularities as the other constants approach their Einsteinian values, in contrast to those with $\lambda=1$ which tend to the known Einstein gravity solutions. At the quantum level, the resulting Wheeler-DeWitt equation is explicitly solved for $\lambda=1,\,\sigma=0$ and $\lambda=\frac{1}{3}$. The ensuing wave-functions diverge in the  Einsteinian limit.
\end{abstract}
%%%%%%%%%%%%%%%%%%%%%%%%%%%%%%%%%%%%%%%%%%%%%%%%%%%%%%%%%%%%%%%%%%%%%%%%%%%%%%%%%%%%%%%%

\newpage
\section{Introduction}
It could be stated that, an important feature of general relativity is the covariance of the theory under four dimensional diffeomorphisms of the spacetime manifold. Recently, a new theory of gravity was proposed, with the aim of being complete in the UV \cite{Hor1}. Its basic assumption is the admittance that the aforementioned covariance is not a fundamental property of the theory, but arises rather accidentally in the context of lower energies. The resulting geometry is then exhibiting a non isotropic time and space scaling invariance
\begin{equation*}
t\mapsto b^z t \;\; , \;\; x^i\mapsto b x^i.
\end{equation*}
As a consequence, one must start from an action functional and equations of motion which involve higher derivatives in the spatial coordinates. On the other hand, it is an advantage that there are no higher derivatives in time. A theory was constructed for various values of the critical exponent $z$ ($z=2$ \cite{Hor2}, $z=3$ and $z=4$ \cite{Hor1}). \\
In the context of this theory one has to consider a four dimensional differentiable manifold, $\mathcal{M}$, with a codimension one foliation $\mathcal{F}$ and admit as the ``gauge invariance" group the foliation preserving diffeomorphisms, i.e. coordinate transformations of the restricted form
\begin{equation} \label{admittedtr}
\tilde{t}=f(t) \;\; , \;\; \tilde{x}^i=g^i(t,x^j).
\end{equation}
With this reasoning the author of \cite{Hor1} modified the 3+1 decomposed Einstein - Hilbert action by adding one extra coupling constant in the kinetic term that destroys four dimensional covariance; As far as the potential part is concerned, extra terms were added under the condition of detailed balance. The latter results in a potential term of the lagrangian density, which up to a coupling constant is
\begin{equation*}
\mathcal{L}_{potential}= E_{ij}G^{ijkl}E_{kl}
\end{equation*}
with $G^{ijkl}$ the generalized Wheeler - DeWitt supermetric and $E_{ij}$ a three tensor that is obtained by a variational principle $\sqrt{g}E^{ij}=\dfrac{\delta W(g_{kl})}{\delta g_{ij}}$ for some action $W$ \cite{Hor1}. \\
There are many papers studying certain cosmological implications of the theory, mostly for a FRW metric, either free of matter or with a scalar field (see for example \cite{KirKor}-\cite{NMRE}, or for an excellent review on Ho\v rava - Lifshitz cosmology, \cite{Mukohyama2} and the references therein). In this paper we are interested in following the point of view of this theory to obtain solutions in the classical context, for a diagonal Bianchi type III model in vacuum, which, as we know, has an anisotropic spatial metric, and compare with analogous solutions from Einstein's equations. Moreover, we proceed with the canonical quantization of the axisymmetric case and derive the Wheeler - DeWitt equation for this cosmological model, giving a solution for specific values of some of the coupling constants. The results obtained are also compared to those obtained by quantizing the Einsteinian action.

\section{Equations of motion}
Our starting point is the action \cite{KirKor}
\begin{eqnarray} \label{action}
S = \int dt d^{3}x \sqrt{g} N \left( \alpha(K_{ij}K^{ij}-\lambda K^{2})+\frac{\beta}{g} C_{pq}C^{pq}+\gamma \frac{\epsilon^{ipk}}{\sqrt{g}}R_{il}R^{l}_{\; k;p}+\zeta R^{i}_{\; j} R^{j}_{\; i}+\eta R^2+\xi R+ \sigma \right)
\end{eqnarray}
where $g$ is the determinant of the three dimensional metric $g_{ij}$, N is the lapse function, $K_{ij}=\frac{1}{2N}(\dot{g}_{ij}-N_{i;j}-N_{j;i})$ the extrinsic curvature corresponding to the spatial metric, $K$ its trace, $C^{pq}=\epsilon^{pkl}(R^{q}_{\; l}-\frac{1}{4}R \delta^{q}_{\; l})_{;k}$ the Cotton - York tensor density and $R_{ij}$, $R$ the Ricci tensor and scalar respectively. Each term in (\ref{action}) comes with its own coupling constant $\alpha$, $\lambda$, $\beta$, $\gamma$, $\zeta$, $\eta$, $\xi$ and $\sigma$. The two parts of the kinetic term are distinguished by the
existence of $\lambda$. For $\lambda=1$ we get the kinetic term of Einstein's theory and the difference is restricted to the extra terms in the potential part and the different coupling constants that bind them together. We note here that ; stands for covariant differentiation with respect to the spatial metric and all the latin indices run from 1 to 3. \\
Variation of the action by $\delta g_{mn}$ results in the following spatial equations of motion
\begin{eqnarray} \nonumber
E^{mn} & \equiv & - \dfrac{\beta N}{2 g}g^{mn}C_{pq}C^{pq} - \beta \dfrac{\epsilon^{qkl}}{2g}(N C_{ql})_{;k}R^{mn}
+\beta \dfrac{\epsilon^{qkl}}{4g}G^{rsmn}(N C_{ql})_{;ksr} - \\ \nonumber
& & - \dfrac{\beta}{2g}\Bigg(N \epsilon^{qkm} C_{q}^{\; l} g^{sn} + N \epsilon^{qkl} C_{q}^{\; m} g^{sn}
-N \epsilon^{qkn} C_{q}^{\; m} g^{sl}-N \epsilon^{qks} C_{q}^{\; l} g^{mn}+(m\leftrightarrow n)\Bigg)_{;ksl}- \\ \nonumber
& & - \dfrac{\beta N}{4g}(\epsilon^{qkm} C_{q}^{\; n} R_{;k}+\epsilon^{qkn} C_{q}^{\; m} R_{;k})+ \\ \nonumber
& & + \dfrac{\beta}{2g}\Bigg(N \epsilon^{qkl} C_{q}^{\; m}R^{n}_{\; l}
+N \epsilon^{qml} C_{q}^{\; k}R^{n}_{\; l}-N \epsilon^{qnl} C_{q}^{\; m}R^{k}_{\; l}+(m\leftrightarrow n)\Bigg)_{;k}- \\ \nonumber
& & -\gamma \dfrac{\epsilon^{ipk}}{2\sqrt{g}}N(R_{i}^{\; m}R^{n}_{\; k;p}+R_{i}^{\;n}R^{m}_{\; k;p})+
\dfrac{\gamma}{4\sqrt{g}}(N R_{qk;p})_{;sl}\Bigg(\epsilon^{mpk}g^{lq}g^{sn}+\epsilon^{lpk}g^{mq}g^{sn}- \\ \nonumber
& &  -\epsilon^{lpk}g^{sq}g^{mn}-\epsilon^{mpk}g^{nq}g^{sl}+(m\leftrightarrow n)\Bigg)- \\ \nonumber
& & - \dfrac{\gamma}{4\sqrt{g}} \Bigg(N \epsilon^{ipl}g^{sn}R_{i}^{\; m} +N \epsilon^{ipm}g^{sn}R_{i}^{\; l}
-N \epsilon^{ips}g^{mn}R_{i}^{\; l}-N \epsilon^{ipn}g^{sl}R_{i}^{\; m}+(m\leftrightarrow n) \Bigg)_{;psl} + \\ \nonumber
& & + \dfrac{\gamma}{4 \sqrt{g}} \Bigg(N \epsilon^{imk} R_{i}^{\; l} R^{n}_{\; k}-N \epsilon^{ink} R_{i}^{\; m} R^{l}_{\; k}+(m\leftrightarrow n) \Bigg)_{;l}+ \\
\nonumber
& & + \dfrac{\zeta N}{2} R^{ij} R_{ij} g^{mn}-2 \zeta N R^{m}_{\; j} R^{jn}+ \\ \nonumber 
& & + \zeta \left((N R^{jm})_{;ij}g^{in}+(N R^{jn})_{;ij}g^{im}-(N R^{mn})_{;ij}g^{ij}-(N R^{ij})_{;ij}g^{mn}\right)+ \\
\nonumber
& &+ \dfrac{\eta N}{2} R^{2} g^{mn}-2 \eta N R R^{mn} + \eta (N R)_{;pq} G^{pqmn}+ \\
\label{eqspa}
& &+ \dfrac{\sigma N}{2} g^{mn}+\dfrac{\xi N}{2} R g^{mn}-\xi N R^{mn}+\dfrac{\xi}{2} N_{;pq}G^{pqmn}=0
\end{eqnarray}
Where $(m\leftrightarrow n)$ stands for the repetition of the terms in the parenthesis followed by an interchange in the m, n indices and $G^{pqmn}=g^{pm}g^{qn}+g^{pn}g^{qm}-2g^{pq}g^{mn}$ is the covariant supermetric. Variation by $\delta N_{i}$ yields the linear constraints
\begin{eqnarray} \label{eqlin}
E^{0i} \equiv 2 \alpha (K^{ij}_{\; ;j}-\lambda g^{ij}K_{;j})=0
\end{eqnarray}
and finally variation by the lapse function N leads to the quadratic constraint
\begin{eqnarray} \label{eqq}
E^{00} \equiv - \alpha (K_{ij}K^{ij} - \lambda K^{2})+\beta C_{ij}C^{ij}+
\gamma \dfrac{\epsilon^{ijk}}{\sqrt{g}}R_{il}R^{l}_{\; k;j}+\zeta R_{ij}R^{ij}+
\eta R^{2} + \xi R +\sigma=0
\end{eqnarray}

\section{Diagonal Bianchi type III, classical case}
As it is known \cite{Chris2001} group automorphisms can be used for simplifying Einstein's equations, since the automorphism generators are Lie point symmetries of the system. In our case we will consider a diagonal scale factor matrix, which, by the procedure described in \cite{typeIII2} and \cite{typeIII}, must be written as
\[\gamma_{\alpha\beta}=
\begin{pmatrix}
e^{u_1(t)+u_3(t)} & 0 & 0 \\
0 & e^{u_2(t)+u_3(t)} & 0 \\
0 & 0 & e^{u_3(t)}
\end{pmatrix}\]
In this particular parametrization of the $\gamma_{\alpha \beta}$'s the automorphisms generators are cast into canonical form, so that the ensuing equations become of first order with respect to the corresponding velocities. The invariant basis 1-forms are (\cite{Ryan})
\[\sigma_i^{\alpha}=
\begin{pmatrix}
0 & e^{-x} & 0 \\
0 & 0 & 1 \\
1 & 0 & 0
\end{pmatrix}\]
while the spatial part of the full metric is given by
\[g_{ij}=\gamma_{\alpha\beta}\sigma^{\alpha}_i \sigma^{\beta}_j=
\begin{pmatrix}
e^{u_3(t)} & 0 & 0 \\
0 & e^{-2x+u_1(t)+u_3(t)} & 0 \\
0 & 0 & e^{u_2(t)+u_3(t)}
\end{pmatrix}\]
One can easily check that this metric corresponds to a three dimensional conformally flat space. This means that the corresponding Cotton - York tensor is zero. Moreover, one can also see that the covariant derivative of the $R_{ij}$ tensor is also zero. Since the invoked coordinate transformations in \cite{Chris2001} are precisely of the form (\ref{admittedtr}) we can, without any loss of generality, take the shift vector $N^i$ to be zero. Then the linear constraint (\ref{eqlin}) yields the equation
\begin{equation} \label{lin}
\frac{\alpha e^{-u_3(t)}\dot{u_1}(t)}{N(t)}=0
\end{equation}
while the quadratic constraint (\ref{eqq}) becomes
\begin{eqnarray} \label{sqcon} \nonumber
 4N(t)^2 \left(2\zeta  +  4\eta +  e^{u_3(t)}(-2\xi+\sigma e^{u_3(t)})\right)+  \alpha  e^{2u_3(t)}  \left( (\lambda-1)(\dot{u_1}(t)^2  +  \dot{u_2}(t)^2) + \right. \\ \left. + (3\lambda-1)(2 \dot{u_2}(t)\dot{u_3}(t)+3\dot{u_3}(t)^2+2\dot{u_1}(t)\dot{u_3}(t))+2\lambda \dot{u_1}(t)\dot{u_2}(t)\right) =0.
\end{eqnarray}
Finally, the three nonzero spatial equations reduce to
\begin{eqnarray} \label{spa1} \nonumber
E_{III}^{11} \equiv -4N^3 (2\zeta+4\eta-\sigma e^{2u_3})+ \alpha e^{2u_3} \left(4\dot{N}\dot{u}_3+N(\dot{u}_1^2+\dot{u}_2^2-  3\dot{u}_3^2-4\ddot{u}_3)\right)+ \\
+ \alpha \lambda e^{2u_3} \left( N (\dot{u}_1+\dot{u}_2+3\dot{u}_3)^2+ 4 (-\dot{N}(\dot{u}_1+\dot{u}_2+3\dot{u}_3)+ N(\ddot{u}_1+\ddot{u}_2+\ddot{u}_3))\right)=0
\end{eqnarray}
\begin{eqnarray} \label{spa2} \nonumber
E_{III}^{22} \equiv -4N^3 (2\zeta+4\eta-\sigma e^{2u_3})-4\alpha e^{2u_3}\dot{N}\left((\lambda-1)\dot{u}_1+\lambda \dot{u}_2+(3\lambda-1)\dot{u}_3 \right) + \\ \nonumber
+ \alpha N e^{2u_3} \left( (\lambda-1)\dot{u}_1^2+(\lambda+1)\dot{u}_2^2+6\lambda\dot{u}_2\dot{u}_3+3(3\lambda-1)\dot{u}_3^2+
2(\lambda-1)\dot{u}_1(\dot{u}_2+3\dot{u}_3) + \right. \\ \left. + 4(\lambda-1)\ddot{u}_1+4\lambda \ddot{u}_2+4(3\lambda-1)\ddot{u}_3\right)=0
\end{eqnarray}
\begin{eqnarray} \label{spa3} \nonumber
E_{III}^{33} \equiv 4N^3 \left(2\zeta+4\eta+e^{u3}(-2\xi+\sigma e^{u_3})\right)- 4\alpha \dot{N}e^{2u_3} \left(\lambda \dot{u}_1+(\lambda-1)\dot{u}_2+(3\lambda-1)\dot{u}_3 \right)+ \\ \nonumber
\alpha N e^{2u_3} \left((\lambda+1)\dot{u}_1^2+(\lambda-1)\dot{u}_2^2+6(\lambda-1)\dot{u}_2\dot{u}_3+ 3(3\lambda-1)\dot{u}_3^2+2\dot{u}_1((\lambda-1)\dot{u}_2+3\lambda\dot{u}_3) + \right. \\ \left.
4\lambda \ddot{u}_1+ 4(\lambda-1)\ddot{u}_2+4(3\lambda-1)\ddot{u}_3 \right)=0
\end{eqnarray}
It is useful to observe that in equations (\ref{spa1}), (\ref{spa2}) and (\ref{spa3}) there is no actual need to take the square root of $N(t)^2$: dividing these equations by $N(t)$ we see that only $N(t)^2$ and $\frac{\dot{N}(t)}{N(t)}\equiv \frac{1}{2 N^2}\frac{d}{dt}(N^2)$ appear. Therefore if we do not ``gauge" fix the lapse a priori, solutions with any signature will be attained. Equation (\ref{lin}) implies that $u_1(t)$ must be a constant. Upon setting $u_1(t)=c_1$, where $c_1$ is a real constant, equations $E_{III}^{11}=0$ and $E_{III}^{22}=0$ become identical. At this point we can solve equation (\ref{sqcon}) with respect to the lapse function and substitute its value into the independent spatial equations (\ref{spa1}) and (\ref{spa3}). This action leads to one final independent equation in terms of the remaining functions $u_2(t)$ and $u_3(t)$. Since this equation is of first order with respect to $\dot{u}_2(t)$ (i.e. $u_2(t)$ does not enter it), while of second order in $u_3(t)$, it is almost mandatory to choose $u_3(t)$ as time. With the specific choice $u_3(t)=\ln(t)$ and upon setting $u_2(t)=\int \omega(t)dt$ we arrive at a final equation of first order in $\omega$:
\begin{eqnarray} \label{abel} \nonumber
\dot{\omega} = &-& \dfrac{t(\lambda-1)(2\lambda\zeta+4\lambda\eta+t(\sigma t-(\lambda+1)\xi))}{2(3\lambda-1)(2\zeta+4\eta+t(\sigma t-2\xi))}\omega^3  -\dfrac{4\eta(3\lambda-1)+3\zeta(3\lambda-1)+t(2\sigma t-\xi-3\lambda\xi)}{2(2\zeta+4\eta+t(\sigma t-2\xi))}\omega^2 \\
&-&\dfrac{18\lambda\zeta+36\lambda\eta+t(5 \sigma t-5\xi-9\lambda\xi)}{2t(2\zeta+4\eta+t(\sigma t-2\xi))}\omega -\dfrac{3(3\lambda-1)(2\zeta+4\eta-\xi t)}{2t^2(2\zeta+4\eta+t(\sigma t-2\xi))}
\end{eqnarray}
This is an Abel equation of the first kind. We did not manage to identify the above equation with any known integrable class, so in order to simplify the situation, in the following subsection we will set $\lambda=1$. \\
But, before we do that, we have to investigate the situation where the coefficient of $N(t)^2$ in (\ref{sqcon}) vanishes, so that this equation does not determine $N(t)$. This happens for
\begin{eqnarray} \label{solden}
u_3(t)= \ln(\dfrac{\xi\pm\sqrt{\xi^2-2\sigma(\zeta+2\eta)}}{\sigma})
\end{eqnarray}
and then the quadratic constraint (\ref{sqcon}) dictates
\begin{eqnarray} \label{sq2}
(\lambda-1)\dot{u}_2(t)=0
\end{eqnarray}
Since the case $\lambda=1$ will be treated in the next subsection, we will here restrict ourselves to the case $u_2(t)=c_2$ with $c_2$ a real constant. With $u_2$ being constant and by the help of (\ref{solden}), equation (\ref{spa3}) is identically satisfied, while the other independent spatial equation (\ref{spa1}) ($\equiv$ (\ref{spa2})) becomes
\begin{eqnarray} \label{eq1}
\left(2(\zeta+2\eta)-\frac{(\xi\pm\sqrt{\xi^2-2\sigma(\zeta+2\eta)})^2}{\sigma}\right)N(t)^3=0
\end{eqnarray}
Since the lapse function cannot be zero, in order for a solution to exist, we have to reduce the number of independent coupling constants. We solve (\ref{eq1}) for $\sigma$ and find that $\sigma=\frac{\xi^2}{2(\zeta+2\eta)}$ for both values of $u_3$. The spatial metric has now become
\[g_{ij}=
\begin{pmatrix}
2\frac{(\zeta+2\eta)}{\xi} & 0 & 0 \\
0 & 2\frac{(\zeta+2\eta)}{\xi}e^{c_1-2x} & 0 \\
0 & 0 & 2\frac{(\zeta+2\eta)}{\xi}e^{c_2}
\end{pmatrix}\]
and one can write the line element as
\begin{eqnarray} \label{lestat}
ds^2=\epsilon dt^2+2\frac{(\zeta+2\eta)}{\xi}dx^2+e^{-2x}dy^2+dz^2
\end{eqnarray}
where $\epsilon=\pm 1$ and the following simplifications have been made: since the lapse has not been defined by (\ref{sqcon}), a reparametrization of time has been used to set $N(t)=\epsilon$ and the spatial coordinates $(y,z)$ have been rescaled in order to absorb the constant factor $\frac{(\zeta+2\eta)}{\xi}$ together with the non essential constants $c_1$ and $c_2$. As we see, only the coupling constants $\xi, \zeta$ and $\eta$ appear in (\ref{lestat}). This is expected by the assumption (\ref{solden}) and its consequences; since the ensuing pseudometric is static $\lambda$ is excluded from it. Moreover, this is a somewhat isolated solution, in the sense that in the Einsteinian limit ($\zeta \rightarrow 0$, $\eta\rightarrow 0$, $\xi \rightarrow \alpha$) it develops a curvature singularity, because the Ricci scalar is $R=-\frac{\xi}{\zeta+2\eta}$.
This particular space admits six killing vectors
\begin{gather*}
\pmb{\xi}_1  = \frac{\partial}{\partial y}, \quad \pmb{\xi}_2 = \frac{\partial}{\partial z}, \quad \pmb{\xi}_3= \frac{\partial}{\partial x} +y \frac{\partial}{\partial y} \\
\pmb{\xi}_4 = y \frac{\partial}{\partial x}+\frac{1}{2}\left(y^2 -e^{2x}\frac{2(\zeta+2\eta)}{\xi} \right)\frac{\partial}{\partial y}, \quad \pmb{\xi}_5=\frac{\partial}{\partial t} \quad \pmb{\xi}_6=z\frac{\partial}{\partial t} +\frac{\partial}{\partial z}
\end{gather*}
The first three are what we would expect from the isometry group and satisfy $\mathcal{L}_{\xi_A}\sigma^{\alpha}_i=0$, with $A=1,2,3$. The fourth depends on the metric components and is of the form investigated in \cite{Szafron} and \cite{Pap}. Usually this kind of vector does not exist as spacetime killing field, but is limited on the spacelike hyper-surface \cite{typeIII}. The non zero structure constants of the algebra closed by these six killing fields are $C_{13}^1=-C_{31}^1=C_{14}^3=-C_{41}^3=C_{26}^5=-C_{62}^5=C_{34}^4=-C_{43}^4=C_{56}^2=-C_{65}^2=1$. Finally, the metric exhibits the property that its Riemmann tensor has vanishing covariant derivative $R_{IJKL|M}=0$ \cite{Coley} (I,J.. run from 0 to 3 while $|$ stands for 4d covariant differentiation). Therefore there are not higher derivative curvatures.

\subsection{Case $\lambda = 1$}
Under the value $\lambda = 1$, as already mentioned, the kinetic term of the theory is identical to the corresponding of Einstein's gravity. By this assumption, equation (\ref{abel}), conveniently becomes a generalized Riccati equation.
\begin{eqnarray} \label{ric}
\dot{\omega} = -\omega^2 -\dfrac{18\zeta+36\eta-14\xi t+5\sigma t}{2t(2\zeta+4\eta+t(\sigma t^2-2\xi))}\omega -\dfrac{3(2\zeta+4\eta-\xi t)}{t^2(2\zeta+4\eta+t(\sigma t-2\xi))}
\end{eqnarray}
whose solution is
\begin{eqnarray}
\omega(t)=\dfrac{3(6c_2 \xi t+12c_2(\zeta+2\eta)+\sqrt{t})}{t(3c_2 \sigma t^2-18c_2(\xi t +\zeta+2\eta)-2\sqrt{t})}
\end{eqnarray}
with $c_2$ being the constant of integration. The lapse function becomes
\begin{equation}
N(t)=3\sqrt{\dfrac{\alpha c_2}{-6c_2\sigma t^2+36c_2(\xi t+\zeta+2\eta)+4\sqrt{t}}}
\end{equation}
while the three dimensional metric is
\[g_{ij}=
\begin{pmatrix}
t & 0 & 0 \\
0 & t e^{c_1-2x} & 0 \\
0 & 0 & \dfrac{3c_2\sigma t^2-18c_2(\xi t+\zeta+2\eta)-2\sqrt{t}}{t}
\end{pmatrix}\]
From the above metric, we see that $c_1$ is not an essential constant, since it can be extinguished by a simple scaling of the $y$ coordinate. The corresponding line element is
\begin{eqnarray} \label{lineric}
ds^2=\dfrac{9\alpha c_2}{2 P(t)}dt^2+tdx^2+ t e^{-2x}dy^2+\dfrac{P(t)}{t}dz^2
\end{eqnarray}
where $P(t)\equiv 3c_2\sigma t^2-18c_2(\xi t+\zeta+2\eta)-2\sqrt{t}$. The integration constant $c_2$ and the positive or negative value of $P(t)$ in some open interval of $t \in \mathbb{R}^+$ determines the signature of the line element: (a) $c_2>0$ and  $P(t)>0$ the signature is Euclidean, (b) $c_2<0$ and  $P(t)>0$ the signature is Lorentzian and (c) $c_2>0$ and  $P(t)<0$ the signature is neutral $(-,+,+,-)$. It is interesting that despite the non linearity of the equations, the above line element can be identified with the known solutions of Einstein gravity: by setting $\zeta=\eta=\sigma=0$ and $\xi=\alpha$ we arrive to vacuum diagonal type III (\cite{Mac}, \cite{typeIII}), while setting $\zeta=\eta=0$ and $\xi=\alpha$ we obtain the cosmological constant type III solutions (\cite{Valent}).  \\
As we mentioned before, we also have to study the case in which (\ref{solden}) holds true. In view of (\ref{solden}), (\ref{sq2}) the quadratic constraint and (\ref{spa3}) become identically zero and we are left with the independent spatial equation
\begin{eqnarray} \nonumber
\alpha \left(\xi^2-\sigma (\zeta+2\eta)+\xi \sqrt{\xi^2-2\sigma (\zeta+2\eta)}\right)\left( N(2 \ddot{u}_2+\dot{u}_2^2)-2\dot{N}\dot{u}_2 \right)+ \\
+ 2\sigma \left(\xi^2-2\sigma (\zeta+2\eta)  +\xi \sqrt{\xi^2-2\sigma (\zeta+2\eta)}\right) N^3=0
\end{eqnarray}
We have the freedom to choose $u_2(t)=t$ and thus the above equation can be integrated for the dependent variable $N(t)$. The result is
\begin{eqnarray}
N(t)^2=\dfrac{e^{t+\mu_1 c_2}}{1-\mu_2 e^{t+\mu_1 c_2}}
\end{eqnarray}
where
\begin{eqnarray} \label{mu1}
\mu_1&=&\left(\dfrac{\xi \pm \sqrt{\xi^2 - 2\sigma (\zeta+2\eta)}}{\sigma}\right)^2 \\ \mu_2&=&-\frac{4\mu_1}{2\alpha}\left(2(\zeta+2\eta)-\mu_1 \sigma\right)
\end{eqnarray}
and $c_2$ is the constant of integration. With these substitutions the spatial metric assumes the form
\[g_{ij}=
\begin{pmatrix}
\sqrt{\mu_1} & 0 & 0 \\
0 & \sqrt{\mu_1} e^{c_1-2x} & 0 \\
0 & 0 & \sqrt{\mu_1} e^t
\end{pmatrix}\]
and the line element can be written as
\begin{eqnarray} \label{le1}
ds^2=-\dfrac{e^{t}}{1-\mu_2 e^{t}}dt^2+\sqrt{\mu_1}dx^2+e^{-2x}dy^2+e^tdz^2
\end{eqnarray}
Again, depending on the sign of $1-\mu_2 e^{t}$ the solution can be both of Lorentzian or Euclidean signature: if $\mu_2$ is negative, then the metric is Lorentzian for all $t\in \mathbb{R}$, whereas if $\mu_2>0$ the solution is Lorentzian for $t<-\ln \mu_2$ and Euclidean for $t>-\ln \mu_2$.  In the Einstenian limit $\xi \rightarrow \alpha$, $\zeta \rightarrow 0$ and $\eta \rightarrow 0$ (we don't set $\sigma \rightarrow 0$ because it obviously leads to singularity), we are led to a singular pseudometic if we choose the minus solution in (\ref{mu1}). On the contrary, for the plus value in (\ref{mu1}) we are left with a pseudometric that is solution of Einstein's equations plus a cosmological constant with $\Lambda=\frac{\sigma}{2\alpha}$. The space-time with line element (\ref{le1}) admits six killing vectors
\begin{gather*}
\pmb{\xi}_1  = \frac{\partial}{\partial y}, \quad \pmb{\xi}_2 = \frac{\partial}{\partial z}, \quad \pmb{\xi}_3= \frac{\partial}{\partial x} +y \frac{\partial}{\partial y} \\
\pmb{\xi}_4 = y \frac{\partial}{\partial x}+\frac{1}{2}\left(y^2 -e^{2x}\sqrt{\mu_1} \right)\frac{\partial}{\partial y},\\
\pmb{\xi}_5=e^{-\frac{z}{2}} \sqrt{e^{-t}-\mu_2}\frac{\partial}{\partial t}+e^{-\frac{z}{2}} \sqrt{e^{-t}-\mu_2}\frac{\partial}{\partial z}, \\
\pmb{\xi}_6=e^{\frac{z}{2}} \sqrt{e^{-t}-\mu_2}\frac{\partial}{\partial t}-e^{\frac{z}{2}} \sqrt{e^{-t}-\mu_2} \frac{\partial}{\partial z}
\end{gather*}
The non zero structure constants of the corresponding algebra are: $C_{13}^1=-C_{31}^1=C_{14}^3=-C_{41}^3=C_{34}^4=-C_{43}^4=1$, $C_{26}^6=-C_{62}^6=C_{52}^5=-C_{25}^5=\frac{1}{2}$, $C_{56}^2=-C_{65}^2=\mu_2$. Again the covariant derivative of the Riemmann tensor vanishes ($R_{IJKL|M}=0$). The constants $\mu_1$, $\mu_2$ appearing in the line element can both be seen to be essential by considering the curvature scalars $R=-\frac{4+\sqrt{\mu_1} \mu_2}{2 \sqrt{\mu_1}}$, $R_{IJKL}R^{IJKL}=\frac{16+\mu_1\mu_2^2}{8\mu_1}$. Since the system can be solved for $\mu_1$, $\mu_2$ in terms of these two curvature scalars, there is no spacetime coordinate transformation that can alter their values; thus they are essential also in the context of the present theory which allows less freedom i.e. transformations (\ref{admittedtr}).

\subsection{Case $\lambda=\frac{1}{3}$}
Equation (\ref{abel}) implies that we also have to distinguish the case $\lambda=\frac{1}{3}$, since $(3\lambda-1)$ appears in the denominator. This is the value for which $Trace(K^i_j-\lambda K \delta^i_j)$ becomes zero. So, for that specific value of $\lambda$, we proceed in the same manner, and after solving (\ref{sqcon}) with respect to the lapse function and substituting in the spatial equations, we are left with the two independent equations produced by (\ref{spa1}) and (\ref{spa3})
\begin{eqnarray} \label{onethird1}
\left( 2(\zeta+2\eta)+e^{u_3}(-4\xi+2\sigma e^{u_3})\right)\dot{u}_2^3(\dot{u}_2+\dot{u}_3)=0 \\ \label{onethird2}
\left( 2(\zeta+2\eta)+e^{u_3}(-4\xi+2\sigma e^{u_3})\right)\dot{u}_2^3\dot{u}_3=0
\end{eqnarray}
When we subtract (\ref{onethird2}) from (\ref{onethird1}) we end up with two possibilities. Either $u_2(t)$ is a constant or $u_3(t)$ is a constant whose value extinguishes the parentheses. It is easy to see that $u_2(t)=constant$ leads, upon setting $\lambda=\frac{1}{3}$, $u_1(t)=constant$ in (\ref{sqcon}), to a zero lapse function. Thus we choose
\begin{eqnarray} \label{13u3}
u_3(t)=\ln\left(\dfrac{2\xi \pm \sqrt{4\xi^2-6\sigma(\zeta+2\eta)}}{3\sigma}\right)
\end{eqnarray}
Upon substitution all the spatial equations are satisfied and for the lapse function we have
\begin{eqnarray}
N(t)^2=\frac{\alpha\left(2\xi\pm\sqrt{4\xi^2-6(\zeta+2\eta)} \right)^2}{12\sigma\left(- \xi^2+6\sigma(\zeta+2\eta)\mp\xi\sqrt{4\xi^2-6\sigma(\zeta+2\eta)}\right)}\dot{u}_2(t)^2
\end{eqnarray}
with $u_2(t)$ an arbitrary function of time. \\
By setting $\nu_1=\frac{2\xi\pm\sqrt{4\xi^2-6\sigma(\zeta+2\eta)}}{3\sigma}$ and $\nu_2=\frac{3 \alpha \sigma}{4\left(- \xi^2+6\sigma(\zeta+2\eta)\mp\xi\sqrt{4\xi^2-6\sigma(\zeta+2\eta)}\right)}$. The spatial metric is written as
\[g_{ij}=
\begin{pmatrix}
\nu_1 & 0 & 0 \\
0 & \nu_1 e^{c_1-2x} & 0 \\
0 & 0 & \nu_1 e^{u_2(t)}
\end{pmatrix}\]
while, upon choosing $u_2(t)=t$, the four dimensional line element is
\begin{eqnarray} \label{le2}
ds^2=-\nu_1^2 \nu_2 dt^2 +\nu_1 dx^2+e^{-2x}dy^2+e^{t}dz^2
\end{eqnarray}
where again the necessary simplifications have been made. Once more, the relation of the values of the various coupling constants determines the signature via the sign of $\nu_2$. The corresponding Riemmann tensor has vanishing covariant derivative and the two constants $\nu_1$, $\nu_2$ are essential, since $R=\frac{1-4\nu_1\nu_2}{2\nu_1^2 \nu_2}$ and $R_{IJKL}R^{IJKL}=\frac{1+16\nu_1^2 \nu_2^2}{8\nu_1^4\nu_2^2}$. At the limit where $\xi\rightarrow\alpha$, $\zeta\rightarrow 0$, $\eta \rightarrow 0$ we are led to a singularity for the minus solution. On the contrary, the plus solution gives a Riemmanian metric which solves Einstein's equations plus a cosmological constant $\Lambda=\frac{3\sigma}{4\alpha}$. The killing vector admitted by (\ref{le2}) are
\begin{gather*}
\pmb{\xi}_1  = \frac{\partial}{\partial y}, \quad \pmb{\xi}_2 = \frac{\partial}{\partial z}, \quad \pmb{\xi}_3= \frac{\partial}{\partial x} +y \frac{\partial}{\partial y} \\
\pmb{\xi}_4 = y \frac{\partial}{\partial x}+\frac{1}{2}\left(y^2 -e^{2x} \nu_1 \right)\frac{\partial}{\partial y},\quad
\pmb{\xi}_5=-4z \frac{\partial}{\partial t}+ (z^2+4\nu_1^2 \nu_2 e^{-t})\frac{\partial}{\partial z},\quad
\pmb{\xi}_6=-2 \frac{\partial}{\partial t} + z \frac{\partial}{\partial z}
\end{gather*}
forming an algebra with the non zero structure constants $C_{13}^1=-C_{31}^1=C_{14}^3=-C_{41}^3=C_{26}^2=-C_{62}^2=C_{34}^4=-C_{43}^4=C_{56}^5=-C_{65}^5=1$ and $C_{25}^6=-C_{52}^6=2$.

\subsection{Case $\lambda \neq 1$, $\zeta=\eta=\sigma=0$}
Finally, we mention two choices for the parameters that lead to another solution. We leave $\lambda$ as an arbitrary constant different from unity and set $\zeta=\eta=\sigma=0$, or $\zeta=-2\eta \; , \; \sigma=0$. Both choices lead to the same differential equation for $u_2(t)$. If we adopt the choice $u_2(t)=\int \frac{\tilde\omega(t)}{t} dt$, we are left with
\begin{eqnarray} \label{secsol}
4 t \dot{\tilde\omega}+\frac{\lambda^2-1}{3\lambda-1}\tilde{\omega}^3+(3\lambda+1)\tilde{\omega}^2+(1+9\lambda)\tilde{\omega}+3(3\lambda-1)=0
\end{eqnarray}
which has the solution
\begin{eqnarray}
\tilde{\omega}=F^{-1}(f(t))
\end{eqnarray}
with
\begin{eqnarray} \nonumber
F(f(t))=-\frac{2\sqrt{2}}{4(\lambda-3)(3\lambda-1)^{1/2}}\tanh^{-1}\left(\frac{3\lambda-1 + (\lambda-1)f(t)}{\sqrt{2(3\lambda-1)}}\right)- \\ \nonumber - \left( \frac{\lambda+1}{4(\lambda-3)(3\lambda-1)} 2 ln\left(-(3\lambda-1)-(\lambda+1)f(t) \right) - \right. \\  \left. -\frac{1}{4(\lambda-3)(3\lambda-1)^{3/2}} ln \left( 3+f(t)(2+f(t))-\lambda (3+f(t))^2 \right) \right)
\end{eqnarray}
where
\begin{eqnarray}
f(t)= c_2- \dfrac{ln (4t (1-3\lambda))}{4(3\lambda-1)}
\end{eqnarray}
and $c_2$ being the constant of integration. \\
At this point we observe, that the above solution depends only on $\lambda$. None of the parameters $\alpha$ or $\xi$ appear in equation (\ref{secsol}). As it seems the declination from general relativity in the kinetic term, produces more dramatic changes in the results. In the case where we restored the original kinetic term with $\lambda=1$ we got a solution which gives the classic one for type III when the other parameters assume the right values. In the latter case we turned off the extra potential terms and kept $\lambda$ as it is, resulting to a totally different solution.

\section{Quantum case for axisymmetric type III}
The Bianchi type III is a class B model.For these models it is known that the Euler-Lagrange equations from the Einstein - Hilbert reduced action with the scale factors taken as the degrees of freedom, are in general not equivalent to the reduced Einstein's equations. Specifically for type III this problem is solved in the diagonal case with $\gamma_{\alpha\beta}=diag(a^2,b^2,c^2)$ by taking the axisymmetric condition $a=c$ (dictated by the linear constraint). So, a first task is to check, if the same holds for the reduced Lagrangian emanating  from (\ref{action}) and the corresponding equations (\ref{sqcon}), (\ref{spa1}), (\ref{spa2}) and (\ref{spa3}). If we insert the scale factor matrix
\[\gamma_{\alpha\beta}=
\begin{pmatrix}
a(t)^2 & 0 & 0 \\
0 & b(t)^2 & 0 \\
0 & 0 & a(t)^2
\end{pmatrix}\]
into action (\ref{action}), we are led to the reduced Lagrangian
\begin{eqnarray} \label{lagr}
L(a,b,N,\dot{a},\dot{b})=2\alpha (1-2\lambda) \frac{b\dot{a}^2}{N}+\alpha (1-\lambda) \frac{a^2\dot{b}^2}{b N}-4\alpha\lambda\frac{a\dot{a}\dot{b}}{N}-2\xi N b+2(\zeta+2\eta)\frac{N b}{a^2}+\sigma N a^2 b
\end{eqnarray}
As it can be straightforwardly verified, the above Lagrangian gives rise to Euler - Lagrange equations equivalent to the equations of motion (\ref{sqcon}), (\ref{spa1})$\equiv$(\ref{spa2}) and (\ref{spa3}) (translated of course by $u_1(t)\equiv 0,\; u_2(t)\equiv 2\ln (\frac{b(t)}{a(t)}), \; u_3(t)\equiv \ln(a(t))$). We now proceed to the Hamiltonian formulation and the subsequent canonical quantization of this system. The conjugate momenta are given as
\begin{eqnarray}
\Pi_N &=& 0 \\
\Pi_a &=& \dfrac{4\alpha((2\lambda-1)b\dot{a}-\lambda a\dot{b})}{N} \\
\Pi_b &=& \dfrac{2\alpha((\lambda-1)a^2 \dot{b}-2\lambda a b \dot{a})}{N b}
\end{eqnarray}
and thus the $\Pi_N\approx 0$ is the primary constraint of the system.
The Dirac - Bergmann algorithm for constrained systems (\cite{Dirac}, \cite{Sund}) leads to the canonical Hamiltonian, $H_c=\dot{a}\Pi_a+\dot{b}\Pi_b-L=N \mathcal{H}_c$, with
\begin{eqnarray} \label{Hcan}
\mathcal{H}_c=\dfrac{\lambda-1}{8\alpha(3\lambda-1)b}\Pi_a^2+\dfrac{(2\lambda-1)b}{4\alpha(3\lambda-1)a^2}\Pi_b^2
-\dfrac{\lambda}{2\alpha(3\lambda-1)a}\Pi_a \Pi_b-\dfrac{2(\zeta+2\eta)-2\xi a^2+\sigma a^4}{a^2}b
\end{eqnarray}
The requirement of preservation in time of the primary constraint indicates that $\mathcal{H}_c\approx 0$ is the secondary constraint. The algorithm thus terminates for $\lambda\neq \frac{1}{3}$ and we are left with these two first class constraints. The case $\lambda=\frac{1}{3}$ will be treated separately. One can read off the supermetric from the kinetic part of $\mathcal{H}_c$:
\[\Gamma^{\mu\nu}=
\begin{pmatrix}
\dfrac{\lambda-1}{4\alpha (3\lambda-1)b} & -\dfrac{\lambda}{2\alpha(3\lambda-1)a} \\
-\dfrac{\lambda}{2\alpha(3\lambda-1)a} & \dfrac{(2\lambda-1)b}{2\alpha (3\lambda-1)a^2} \\
\end{pmatrix}\]
To canonically quantize the system in the Schr\"{o}dinger picture we adopt the usual choice of basic operators ($\hbar=1$): $\Pi_N  \rightarrow  \hat{\Pi}_N=-i\frac{\partial}{\partial N}$, $\Pi_a  \rightarrow  \hat{\Pi}_a=-i\frac{\partial}{\partial a}$, $\Pi_b  \rightarrow  \hat{\Pi}_b=-i\frac{\partial}{\partial b}$ with associated commutation relations $[N,\hat{\Pi}_N^\prime]  =i  \delta_{N N^\prime}$,$[a,\hat{\Pi}_a^\prime]  =i  \delta_{a a^\prime}$, $[b,\hat{\Pi}_b^\prime]  =i  \delta_{b b^\prime}$. \\
We then follow Dirac's proposal and define the wave function to be annihilated by the quantum analogue of the two first class constraints. The primary $\hat{\Pi}_N \Psi(N,a,b)=0$ informs as that $\Psi$ does not depend on $N$. For the secondary we choose the factor ordering so that the kinetic part of $\hat{\mathcal{H}}_c$ becomes the Laplace - Beltrami operator
\begin{equation}
\hat{\mathcal{H}}_{kin}=\frac{1}{2}\Gamma^{-1/2}\hat{\Pi}_\mu \Gamma^{1/2}\Gamma^{\mu\nu}\hat{\Pi}_\nu
\end{equation}
The Wheeler - DeWitt equation $\hat{\mathcal{H}}_c \Psi (a,b)=0$ becomes
\begin{eqnarray} \label{wdwbig}
\dfrac{\lambda-1}{8\alpha(3\lambda-1)b}\frac{\partial^2 \Psi}{\partial a^2}+\dfrac{(2\lambda-1)b}{4\alpha(3\lambda-1)a^2}\frac{\partial^2 \Psi}{\partial b^2}-\dfrac{\lambda}{2\alpha (3\lambda-1)a}\frac{\partial^2 \Psi}{\partial a\partial b} + \\
+\dfrac{\lambda-1}{8\alpha (3\lambda-1)a b}\frac{\partial\Psi}{\partial a}+\dfrac{2\lambda-1}{4\alpha (3\lambda-1)a^2}\frac{\partial\Psi}{\partial b}+\dfrac{2(\zeta+2\eta)-2\xi a^2+\sigma a^4}{a^2}b=0
\end{eqnarray}
The space of solutions to this partial differential equation is difficult enough to be found in full generality (i.e. for arbitrary values of the parameters). Thus, we give below explicit solutions for $\lambda=1, \; \sigma=0$ and $\lambda=\frac{1}{3}$.

\subsection{Case $\lambda=1$}
Under the assumption $\lambda=1$ (from now on we take $\alpha=1$) equation (\ref{wdwbig}) reduces to
\begin{eqnarray}
\frac{b}{8 a^2}\frac{\partial^2 \Psi}{\partial b^2}-\frac{1}{4 a}\frac{\partial^2 \Psi}{\partial a\partial b}+\frac{1}{8 a^2}\frac{\partial \Psi}{\partial b}+\dfrac{2(\zeta+2\eta)-2\xi a^2+\sigma a^4}{a^2}b=0
\end{eqnarray}
Under the change of variables $(a,b)\mapsto(x,y)$ where $x=a b$ and $y=\frac{1}{b}$ we can write the previous equation without a mixed derivative term. A further simplification by dropping the cosmological term, i.e. setting $\sigma=0$, results in the Wheeler - DeWitt equation being solvable through separation of variables: $\Psi(x,y)\equiv X(x)Y(y)$ whith the two one variable functions satisfy the equations
\begin{eqnarray} \label{sep1}
\dfrac{x^2 X^{\prime\prime}(x)}{8X(x)}+\dfrac{x X^{\prime}(x)}{8 X(x)}+2\xi x^2=2 c^2 \\ \label{sep2}
\dfrac{y^2 Y^{\prime\prime}(y)}{8Y(y)}+\dfrac{y Y^{\prime}(y)}{8 Y(y)}+ \frac{2(\zeta+2\eta)}{y^2}=2 c^2
\end{eqnarray}
with c being an arbitrary constant and the primes indicating differentiation with respect to the arguments. The solutions to these equations are
\begin{eqnarray}
X(x)&=&\kappa_1 J_{4c}(4\sqrt{\xi} x)+\kappa_2 Y_{4c}(4\sqrt{\xi} x) \\
Y(y)&=&\kappa_3 (-1)^{-2c}\Gamma(1-4c) I_{-4c}(4\frac{\sqrt{-\zeta-2\eta}}{y})+ \kappa_4 (-1)^{2c} \Gamma(1+4c) I_{4c}(4\frac{\sqrt{-\zeta-2\eta}}{y})
\end{eqnarray}
where $\kappa_i$, $i=1,2,3,4$ are the constants of integration $J_\nu(x)$, $Y_\nu(x)$ and $I_\nu(x)$ are the Bessel J, Y and I functions respectively, and lastly $\Gamma(x)$ stands for the Gamma function. Translating this result to the old coordinates the solution of the Wheeler - DeWitt without $\sigma$ and for $\lambda=1$ is
\begin{eqnarray} \nonumber
\Psi(a,b) = \left(\kappa_1 J_{4c}(4\sqrt{\xi} a b)+\kappa_2 Y_{4c}(4\sqrt{\xi} ab)\right)
\left(\kappa_3 (-1)^{-2c}\Gamma(1-4c) I_{-4c}(4 b \sqrt{-\zeta-2\eta})+\right. \\ \label{psi1} \left.  + \kappa_4 (-1)^{2c} \Gamma(1+4c) I_{4c}(4 b \sqrt{-\zeta-2\eta}) \right)
\end{eqnarray}
If we want to compare with the results from Einstein's gravity, we must start from the reduced Lagrangian
\begin{eqnarray}
L_E=-\frac{2b \dot{a}^2}{N}-\frac{4 a \dot{a} \dot{b}}{N}-2 N b
\end{eqnarray}
We follow exactly the same procedure, which results in the following Wheeler - DeWitt equation:
\begin{eqnarray} \label{wdwe}
-\frac{b}{8 a^2}\dfrac{\partial^2 \Psi_E}{\partial b^2}+\frac{1}{4 a}\dfrac{\partial^2 \Psi_E}{\partial a \partial b}- \frac{1}{8 a^2}\dfrac{\partial \Psi_E}{\partial b}+2 b \Psi_E=0
\end{eqnarray}
Again, the choice of variables $x=ab$ and $y=\frac{1}{b}$, leads to the extinction of the mixed derivative and allows (\ref{wdwe}) to be solved with a separation of variables. Upon setting $\Psi(x,y)=X(x)Y(y)$ we get:
\begin{eqnarray} \label{sep11}
\dfrac{x^2 X_E^{\prime\prime}(x)}{8X_E(x)}+\dfrac{x X_E^{\prime}(x)}{8 X_E(x)}+2 x^2=2 c^2 \\ \label{sep22}
\dfrac{y^2 Y_E^{\prime\prime}(y)}{8Y_E(y)}+\dfrac{y Y_E^{\prime}(y)}{8 Y_E(y)}=2 c^2
\end{eqnarray}
These are the same equations one would acquire by setting the values $\xi=1, \zeta=\eta=0$ or $\xi=1, \zeta=-2\eta$ in (\ref{sep1})-(\ref{sep2}). The solution of (\ref{wdwe}) is:
\begin{eqnarray}
\Psi_E(a,b) = \left(\kappa_1 J_{4c}(4 a b)+\kappa_2 Y_{4c}(4 ab)\right)
\left(\kappa_3 \cosh (4c \ln(\frac{1}{b}))+i \; \kappa_4 \sinh(4c \ln(\frac{1}{b})) \right)
\end{eqnarray}
The first parenthesis is the same with the one from the first solution under the condition $\xi=1$. The terms in the second parenthesis are now written in the form of ordinary functions, as opposed to the previous case. If we now set $\xi=1, \zeta=\eta=0$ in (\ref{psi1}), we get
\[
  \Psi(a,b) = \left\{
  \begin{array}{l l}
    (\kappa_2+\kappa_3)\left(\kappa_1 J_{0}(4 a b)+\kappa_2 Y_{0}(4 ab)\right) & \quad \text{if $c=0$} \\
    \tilde{\infty} & \quad \text{if $|4c| \in \mathbb{R}^+$}
  \end{array} \right.
\]
while at the same time for $c\rightarrow 0$
\begin{eqnarray} \label{psiezero}
\Psi_{E\; (c\rightarrow 0)}(a,b) = \kappa_3 \left(\kappa_1 J_{0}(4 a b)+\kappa_2 Y_{0}(4 a b)\right)
\end{eqnarray}
We have to note here that (\ref{psiezero}) is not produced by the solution of the set of equations (\ref{sep11}) and (\ref{sep22}) with $c=0$. We just refer to the limit of their solutions with $c\neq 0$ and as $c \rightarrow 0$. So under the limiting case, where the extra coupling constants tend to take values that correspond to Einstein's gravity, we see that $\Psi(a,b)$ diverges, unless we make the specific choice in (\ref{sep1}) and (\ref{sep2}) to set $c=0$. Then the solution becomes essentially the same with $\lim_{c\to 0}\Psi_E(a,b)$, which of course is not the general solution of (\ref{sep11}) and (\ref{sep22}) for $c=0$. The latter is $\Psi_{E\; (c=0)}(a,b) = (\kappa_3+\kappa_4 \ln{\frac{1}{b}}) \left(\kappa_1 J_{0}(4 a b)+\kappa_2 Y_{0}(4 ab)\right)$. \\
One could remove the divergence of $\Psi(a,b)$ as $\zeta$ and $\eta$ tend to zero and $c\neq 0$, by extinguishing one of the constants of integration $\kappa_3$ or $\kappa_4$ (the choice depends on whether the admissible values of c are positive or negative), but that would eventually lead to a zero wave function.

\subsection{Case $\lambda=\frac{1}{3}$}
Equation (\ref{Hcan}) entails the need to treat the case $\lambda=\frac{1}{3}$ separately. If we set this value for $\lambda$ in (\ref{lagr}) and calculate the conjugate momenta $\Pi_a$ and $\Pi_b$ we see that they are not independent, a fact which results in the existence of another constraint. It is easy to check that in our case
\begin{eqnarray}
a \Pi_a + b\Pi_b=0
\end{eqnarray}
or re-writing it in operator form
\begin{eqnarray}
a \frac{\partial}{\partial a} + b\frac{\partial}{\partial b}=0
\end{eqnarray}
An invariant function for this operator is $\frac{b}{a}$, so if we set this fraction as a new variable, the operator will be cast into canonical form, and the transformed Lagrangian will depend on only one velocity. We choose the set of new variables to be $(u,v)$ with $u=\frac{b}{a}$ and $v=a$. Under this transformation and for the specific value $\lambda=\frac{1}{3}$ Lagrangian (\ref{lagr}) becomes
\begin{eqnarray}
L(u,a,N,\dot{u})=\frac{2 a^3 \dot{u}^2}{3Nu}+ N\left(2(\zeta+2\eta)\frac{u}{a}-2\xi u a+\sigma u a^3\right)
\end{eqnarray}
The conjugate momenta are
\begin{eqnarray}
\Pi_N &=& 0 \\
\Pi_a &=& 0 \\
\Pi_u &=& \frac{4 a^3 \dot{u}}{3Nu}
\end{eqnarray}
with $\Pi_N$ and $\Pi_a$ being the primary constraints. Now, we can write the canonical Hamiltonian $H_c=\dot{u}\Pi_u-L$, which turns out to be
\begin{eqnarray}
H_c=N \mathcal{H}_c=N u \left(\frac{3}{8 a^3}\Pi_u^2 -2\frac{(\zeta+2\eta)}{a}+2\xi a-\sigma a^3\right)
\end{eqnarray}
while the total Hamiltonian defined in the full phase space is
\begin{eqnarray} \label{totH}
H_T = H_c + \mu_1 \Pi_N + \mu_2 \Pi_a
\end{eqnarray}
The condition that the constraints must be preserved in time, $\{\Pi_N,H_T\}\approx0$ and $\{\Pi_a,H_T\}\approx0$ leads to the secondary constraints
\begin{eqnarray}
\chi_1 &=& 16  (\zeta +2 \eta ) a^2-16  \xi  a^4+8  \sigma  a^6-3 \Pi_u^2 \\
\chi_2 &=& 2 (\zeta +2 \eta) -4 a^2 \xi +3 a^4 \sigma
\end{eqnarray}
As for their preservation in time, the Poisson bracket of $\chi_1$ with the total Hamiltonian is weakly zero, while $\{\chi_2,H_T\}\approx0$ imposes the condition $\mu_2\approx 0$ in (\ref{totH}). So we have a total of four constraints with $\Pi_N$, $\chi_1$ being first class and $\Pi_a$, $\chi_2$ being second class. We use the definition of the Dirac bracket $\{\;,\;\}_D = \{\;,\;\}-\{\;,\xi_\mu\}\Delta^{-1}_{\mu\nu}\{\xi_\nu,\;\}$ with $\xi_\mu$ being the elements of the set of the second class constraints and $\Delta_{\mu\nu}=\{\xi_\mu,\xi_\nu\}$ the non singular matrix formed by their Poisson brackets. \\
The use of the Dirac brackets allows us to set the second class constraints strongly equal to zero. The basic canonical commutation relations become
\begin{eqnarray} \nonumber
\{N,\Pi_N\}_D &=& 1 \\ \nonumber
\{u,\Pi_u\}_D &=& 1 \\ \nonumber
\{a,\Pi_a\}_D &=& 0
\end{eqnarray}
with the last of them simply emphasizing the redundancy of the corresponding pair of variables. Thus on the further restricted phase space $\chi_2 \simeq 0$ leads to a constant scale factor
\begin{eqnarray} \label{conchi2}
a^2 \approx \frac{2 \xi \pm \sqrt{4 \xi ^2-6 (\zeta +2 \eta ) \sigma }}{3 \sigma }
\end{eqnarray}
It is interesting to note that, as expected from the equivalence between the Hamiltonian and the Lagrangian formulation for constrained dynamics, ($\ref{conchi2}$) is identical to the value of the square of the constant scale factor in the classical solution, as obtained in $(\ref{13u3})$.
Again we choose the operators
\begin{eqnarray*}
\Pi_N & \rightarrow & \hat{\Pi}_N=-i\frac{\partial}{\partial N} \\
\Pi_u & \rightarrow & \hat{\Pi}_u=-i\frac{\partial}{\partial u}
\end{eqnarray*}
The quantum analogue of the primary constraint $\hat{\Pi}_N \Psi(N,u) = 0$ dictates a solution of the form $\Psi=\Psi(u)$. Our Wheeler - DeWitt equation results from $\hat{\chi}_1 \Psi(u)=0$ under the condition (\ref{conchi2}),
\begin{eqnarray}
32\left(4 \xi ^3-9 (\zeta +2 \eta ) \xi  \sigma \pm 2 \xi ^2 \sqrt{4 \xi ^2-6 (\zeta +2 \eta ) \sigma }\mp 3 (\zeta +2 \eta ) \sigma  \sqrt{4 \xi ^2-6 (\zeta +2 \eta ) \sigma }\right)\Psi (u)-81 \sigma ^2\Psi ''(u)=0
\end{eqnarray}
its solution is
\begin{eqnarray}
\Psi (u)= c_1 e^{\sqrt{\kappa}u} + c_2 e^{-\sqrt{\kappa}u}
\end{eqnarray}
with
\begin{eqnarray}
\kappa =\frac{32\left(4 \xi ^3-9 (\zeta +2 \eta ) \xi  \sigma \pm 2 \xi ^2 \sqrt{4 \xi ^2-6 (\zeta +2 \eta ) \sigma } \mp 3 (\zeta +2 \eta ) \sigma  \sqrt{4 \xi ^2-6 (\zeta +2 \eta ) \sigma }\right)}{81 \sigma ^2}
\end{eqnarray}
Upon setting $\zeta=-2\eta$ and $\xi=1$, for the plus value of (\ref{conchi2}), we are led to $\kappa=\frac{256}{81\sigma^2}$. On the other hand, the minus value gives $\kappa=0$, which induces a constant wave function for every value of u.

\section{Discussion}
We have treated a diagonal Bianchi type III cosmological model within Ho\v rava - Lifshitz theory classically and quantum mechanically. \\
At the classical level, the main tools for investigating the solution space were: (a) the automorphisms of the corresponding Lie algebra whose generators constitute Lie point symmetries of the equations of motion and (b) the use of the quadratic constraint as an algebraic equation determining the lapse $N(t)$ and the subsequent replacement of its value in the spatial equations. The virtue of this is twofold: On the one hand it becomes possible to reduce the order of the system of differential equations and ultimately arrive at (\ref{abel}) and the main solution (\ref{lineric}), and on the other hand various signature solutions are attained. When the lapse is not determined by the quadratic constraint, we are led to the marginal metrics (\ref{lestat}) and (\ref{le1}). These metrics are interesting enough, since they are curvature homogeneous spacetimes. Additionally, one might consider (\ref{le1}) for $\mu_2=0$ as equivalent to (\ref{lestat}), since the curvature invariant relations are then identical. But this is not true, because the transformation needed for this identification involves mixing of $t$,$z$ as it can be seen by the corresponding killing vectors. Thus the metrics are not equivalent in the context of Ho\v rava - Lifshitz theory. For the same reason line element (\ref{le1}) can not be considered as static. It is also interesting that for $\lambda=\frac{1}{3}$ the resulting line element (\ref{le2}) is also a curvature homogeneous space despite the fact that in this case the lapse has been determined. For these three spacetimes a noteworthy segregation of the original coupling constants occurs: if we change their values in a way that does not affect the combinations entering the essential constants the classical solutions are not affected, leading one to consider all these models as equivalent.\\
Another important observation concerns the coupling constant $\lambda$. Its value significantly changes the essence of the classical theory. For $\lambda=1$, the general solution (\ref{lineric}) can be considered as the original solution of Einstein's equation plus higher term ``corrections". For $\lambda=\frac{1}{3}$ the solution (\ref{le2}) develops for $\sigma=0$ curvature singularities at the Einsteinian limit, or becomes a solution of Einstein's equation plus a cosmological constant term. The only case for which $\lambda$ becomes utterly unimportant is the static pseudometric (\ref{lestat}), a thing expected since $\lambda$ appears inside the kinetic term. Lastly, it is fortunate that, even though we consider a class B model for which we can write a valid Lagrangian only under the axisymmetric condition, we do not lose this good property for the action of Ho\v ravas' theory. Thus the Euler - Lagrange equations of the reduced action, are identical to the reduced equations of motion.  \\
At the quantum level, we were able to solve the Wheeler - DeWitt equation for the values $\lambda=1$ and $\lambda=\frac{1}{3}$. Even in the case of $\lambda=1$ the Einsteinian limit of the wave function can not reproduce the corresponding Einsteinian wave function, because it diverges. The only non divergent case is when the separation constant is set to zero. \\
Of course, the renouncement of four dimensional covariance is not a problem free choice: it has been observed (\cite{Lu}, \cite{Mukohyama2}, \cite{pro1}) that there are not only difficulties attaining general relativity in the Einsteinian limit ($\lambda\rightarrow 1$) but also stability problems regarding the UV region. Recently, a new approach is proposed in \cite{Hor3}, in an attempt for the theory to overcome its problems. Since we could not obtain the general solution of the Abel equation (\ref{abel}), we can not comment on this issue.

\end{document}